\documentclass[11pt]{article}
 
 \usepackage{amsfonts,amssymb,amsmath}
\usepackage[latin1]{inputenc}

\newcommand{\oo}[0]{\otimes}

\newcommand{\m}[0]{\eta}

\def\s{w}
\def\1{\'{\i}}                           
                           
\def\be{\begin{equation}}
\def\ee{\end{equation}}
\def\bea{\begin{eqnarray}}
\def\eea{\end{eqnarray}}
\def\Drinfel'd double{{\rm d}}

\def\Drinfel'd double{{\rm d}}

\def\>#1{{\bf #1}}

\def\s{s}

 \def\xx{{\xi}}
 
 \def\m{{\eta}}

 \def\xx{{\xi}}

\def\s{w}
                           
\def\Drinfel'd double{{\rm d}}

\def\vv{{\theta}}

\def\C{{\Upsilon}}

\begin{document}

\ \bigskip\bigskip

\begin{center}
\baselineskip 24 pt {\Large \bf  
A (2+1) non-commutative Drinfel'd double spacetime\\ with  cosmological constant}

\end{center}

 \medskip

\begin{center}

{\sc Angel  Ballesteros$^1$, Francisco J. Herranz$^1$ and
Catherine Meusburger$^2$}

{$^1$ Departamento de F\1sica, Universidad de Burgos, 
E-09001 Burgos, Spain}

{$^2$  Department Mathematik,  FAU Erlangen-N\"urnberg, Cauerstr.~11, D-91058 Erlangen, Germany
}
 
e-mail: {angelb@ubu.es, fjherranz@ubu.es,  catherine.meusburger@math.uni-erlangen.de}

\end{center}

\begin{abstract}
We show that the Drinfel'd double associated to the standard quantum deformation $sl_\m(2,\mathbb{R})$ is isomorphic to the (2+1)-dimensional  AdS  algebra
with the initial deformation parameter $\m$ related to the cosmological constant $\Lambda=-\eta^2$.  This gives rise to a generalisation of  
a non-commutative Minkowski spacetime that arises as a consequence of the quantum double symmetry of (2+1) gravity to non-vanishing cosmological constant.
The properties of the AdS quantum double that generalises this symmetry to the  case $\Lambda\neq 0$ are sketched, and it is shown that the new non-commutative AdS spacetime is a nonlinear $\Lambda$-deformation of the Minkowskian one.
\end{abstract}

\medskip 




\section{Introduction}

Following the pioneering work by Snyder~\cite{Snyder}, non-commutative spacetimes have been pursued as an algebraic approach to model  properties of spacetimes that should arise at the Planck scale  
(see, for instance,~\cite{Garay} and references therein).  Essentially, if spacetime coordinates are converted into non-commuting operators in a consistent way (that usually requests invariance properties under certain spacetime transformations) then the discreteness of  the spectrum of spacetime operators~\cite{Snyder}  or the non-trivial commutation relations between them~\cite{DFR} provide a useful description of the expected discretisation or fuzziness of spacetimes in the Planck regime.

In this context, quantum groups~\cite{CP,majid} have provided a mathematically consistent and powerful approach to the rigorous definition of non-commutative spacetimes. In fact, any quantum (Hopf algebra) deformation of a given Lie algebra is associated in a canonical way with a Poisson--Lie (PL) structure  
on the associated Lie group~\cite{Drinfelda,Drib}, and the quantisation of this PL algebra defines a non-commutative algebra of local group coordinates. In particular, if the Lie group  under consideration is a   group of isometries of a certain spacetime (for instance, the Poincar\'e group in the case of Minkowski space), the associated non-commutative spacetime is defined by the commutation rules among the ``quantum" space and time coordinate functions. In this way, different non-commutative algebras associated with Minkowski space have been considered in the literature (see~\cite{kMinkowski,nullplane, RossanoPLB,ktwist} and references therein). However, explicit proposals concerning non-commutative spacetimes with non-vanishing cosmological constant are ---to the best of our knowledge--- still lacking, with the exception of the $\kappa$-AdS space introduced in~\cite{BHBruno} and further studied in~\cite{Marciano}. In this respect, if astrophysical and cosmological tests of Planck-scale phenomena are devised (see~\cite{AmelinoLRR} and references therein), the explicit introduction of the cosmological constant will be essential in order to model the interplay between Planck-scale effects and spacetime curvature.

Moreover, for a given Lie group of isometries there are many possible quantum deformations, and a clear connection between  specific  deformations and any fundamental properties of (quantum) gravity on the corresponding spacetime remains as an important open problem. In (2+1) gravity, the question which  quantum deformations are suitable quantum symmetries for gravity  is easier to address and may provide important insights for higher dimensions. 

This is due to the fact that quantum group symmetries in (2+1)-gravity are not introduced ad hoc or from phenomenological considerations, but can be derived from the classical theory. They arise as the quantum  counterparts of certain PL symmetries that describe the Poisson structure on the phase space of the theory in its formulation as a Chern--Simons (CS) gauge theory.  There is good evidence that the relevant quantum group symmetries are Drinfel'd doubles \cite{cm2,MN,TVir, BK},
  and specific 
 quantum deformations of the corresponding isometry groups were proposed in~\cite{BHMplb}.  Following this approach, the full classification and explicit construction of all the possible Drinfel'd double quantum deformations of the de Sitter (dS) and anti-de Sitter (AdS) groups in (2+1) dimensions that are compatible with the CS formulation of (2+1) gravity was recently given in~\cite{BHMCQG}.

In this article, we derive first results about  the non-commutative spacetimes that arise from these Drinfel'd double symmetries in (2+1)-gravity. In particular, 
 we summarise the main properties of  the ${\bf AdS}_\xi^{2+1}$ quantum spacetime that results from a certain  Drinfel'd double quantum deformation of the isometry group of  the AdS space  that was studied in~\cite{BHMCQG}. By construction, this new AdS non-commutative spacetime  should be connected with (2+1)-gravity, and we show that this is indeed the case: the ${\bf AdS}_\xi^{2+1}$ algebra turns out to be  generalisation to non-vanishing cosmological constant of the $so(2,1)$-non-commutative (2+1) Minkowski spacetime ${\bf M}_\xi^{2+1}$, which  arises naturally  in the quantisation of (2+1)-gravity when the  Poincar\'e group in (2+1) dimensions is considered as the Drinfel'd double  $D(sl(2,\mathbb{R}))$~\cite{Bais, BMS, MW} (see also~\cite{BatistaMajid, Majidtime, Noui} for the Euclidean case based on $D(su(2))$). In particular, ${\bf AdS}_\xi^{2+1}$ is found to be a deformation of the Lie algebra $so(2,1)$,  in which the deformation parameter is directly related with $\Lambda$. Therefore, the quantum group framework here presented shows that a non-vanishing cosmological constant leads to a nonlinear generalization of the previously considered Lie algebraic noncommutative spacetimes.

The article is organised as follows. The next section summarises the construction of non-commutative spacetimes from  Drinfel'd doubles.
This approach is illustrated in section 3,  on the example of the  Poincar\'e  algebra in (2+1) dimensions, which  is the Drinfel'd double $D(sl(2,\mathbb{R}))$  associated to the trivial ({\em i.e.}, non-deformed) Hopf algebra structure of the universal enveloping algebra $U(sl(2,\mathbb{R}))$.  Moreover, it is shown that  the associated non-commutative  Minkowski spacetime ${\bf M}_\xi^{2+1}$  is the quantisation of the PL structure defined by the canonical classical $r$-matrix of  $D(sl(2,\mathbb{R}))$. In section 4, we perform the same construction but taking instead as the  starting point  the standard or  Drinfel'd--Jimbo quantum deformation $U_\m(sl(2,\mathbb{R}))$~\cite{Drinfelda,Jimbo}. In this case the Drinfel'd double associated to this $\m$-deformed Hopf algebra turns out to be the (2+1) AdS algebra, in which the cosmological constant $\Lambda$ that determines the curvature is given by the quantum $sl(2,\mathbb{R})$ deformation parameter as $\Lambda=-\m^2$.  Consequently, the associated non-commutative AdS spacetime ${\bf AdS}_\xi^{2+1}$ is obtained as the quantisation of the PL structure defined by the canonical classical $r$-matrix provided by $D(sl_\m(2,\mathbb{R}))$. 

Section 5 contains the explicit construction of the non-commutative spacetime ${\bf AdS}_\xi^{2+1}$. We first introduce the vector model of the classical isometry group of the (2+1) AdS space and explicitly compute the full  PL structure determined by the canonical  classical $r$-matrix. In particular, we determine the Poisson brackets of  the AdS spacetime coordinates, which are  the classical counterparts of the commutators of spacetime operators in the non-commutative AdS spacetime. The resulting quantum AdS spacetime turns out to be a deformation of the $so(2,1)$ non-commutative spacetime ${\bf M}_\xi^{2+1}$ with the cosmological constant as a deformation parameter. We emphasise that this possibility,  that allows one to to perform a ``cosmological limit'' $\Lambda\to 0$, is a general feature of our construction. In all expressions for ${\bf AdS}_\xi^{2+1}$ and the corresponding quantum algebra,
 the cosmological constant is contained explicitly as a parameter, and the corresponding expressions for ${\bf M}_\xi^{2+1}$ and the corresponding deformation of the Poincar\'e algebra can be recovered as a limit $\Lambda\to 0$. The final section contains some remarks and open problems for future research.


\section{Non-commutative spacetimes from Drinfel'd Doubles}

Let $G$ be  a finite-dimensional Lie group $G$ with  Lie algebra $\mathfrak{g}$ and  $\{Y_i\}$ a basis of $\mathfrak g$. Consider 
 the natural Hopf algebra structure of the universal enveloping algebra $U(\mathfrak{g})$ given by the primitive coproduct map $\Delta_0: U(\mathfrak g)\to U(\mathfrak g)\otimes U(\mathfrak g)$ defined by
\be
\Delta_0(Y)= Y \otimes 1 + 1\otimes Y ,
\qquad 
\forall Y\in \mathfrak{g}.
\label{delta0}
\ee
A quantum algebra $(U_\m(\mathfrak{g}),\Delta_\m)$ is a Hopf algebra deformation of $(U(\mathfrak{g}),\Delta_0)$ in which  the new ``quantum"  coassociative coproduct map $\Delta_\m$ is constructed as a formal power series in the quantum deformation parameter $\m$ as
\be
\Delta_\m=\sum_{k=0}^{\infty} \m^k \Delta_{k}= \Delta_0 + \m\,\Delta_1 + o[\m^2],
\label{defcop}
\ee
and the product of $U(\mathfrak g)$ is modified in such a way  that the deformed coproduct  $\Delta_\m$ becomes an algebra homomorphism $\Delta_\m: U_\m(\mathfrak g)\to U_\m(\mathfrak g)\otimes U_\m(\mathfrak g)$. As the structures are defined as a formal power series in the deformation parameter $\m$, each quantum
deformation  $(U_\m(\mathfrak{g}),\Delta_\m)$ defines  a unique  Lie bialgebra structure $(\mathfrak{g},\delta)$ obtained as the linearisation of $(U_\m(\mathfrak{g}),\Delta_\m)$ in the parameter $\m$. Then, the skew-symmetric part of the first-order deformation $\Delta_1$ of the quantum coproduct $\Delta_\m$~\eqref{defcop} defines the cocommutator map
 $\delta: \mathfrak g\to\mathfrak g\oo \mathfrak g$
\be
\delta(Y_n)=f^{lm}_n\,Y_l\wedge Y_m
\label{cocof}
\ee
and the skew-symmetrisation of the multiplication map gives rise to the usual Lie bracket on $\mathfrak g$.  

In this framework, the Hopf algebra dual to $(U_\m(\mathfrak{g}),\Delta_\m)$ can be interpreted as the non-commutative Hopf algebra of functions on the quantum group $\mbox{Fun}_\m(G)$. The algebra  of  quantum coordinate operators is  obtained
via  Hopf algebra duality from the  deformed coproduct~\eqref{defcop}, and its non-commutativity is a consequence of the fact that  the deformed coproduct~\eqref{defcop} is in general non-cocommutative.  Moreover, the resulting quantum group is  the quantisation of the unique PL  structure on $G$ that is associated to the cocommutator map $\delta$~(\ref{cocof}). In particular, if $G$ is the group of isometries of a spacetime, a given quantum deformation of its Lie algebra will induce a specific non-commutative spacetime.

From this, it could be concluded that non-deformed Hopf algebras would be trivial from the point of view of non-commutative spacetimes. However, this is not  true: the Hopf algebra $(U(\mathfrak{g}),\Delta_0)$ with a trivial coproduct induces interesting non-commutative geometry structures on the double Lie group $D(G)$, whose Lie algebra is the so-called Drinfel'd double Lie algebra $D ({\mathfrak{g}})$ (see~\cite{Bais, BMS,BatistaMajid}). 
This statement can be made explicit by considering
the ``trivial"  Lie bialgebra structure $(\mathfrak{g},\delta_0)$ that corresponds to the ``non-deformed" quantum universal enveloping algebra   $(U(\mathfrak{g}),\Delta_0)$ and is given by
\be
[Y_i,Y_j]= c^k_{ij}Y_k, \qquad  \delta_0(Y)=0,
\label{lib}
\ee
where $c^k_{ij}$ denote the structure constants of $\mathfrak g$ with respect to the basis $\{Y_i\}$.
If we fix a basis $\{y^i\}$ of the vector space $\mathfrak{g} ^*$ dual to $\{Y_i\}$, this yields a pairing
\be
\langle Y_i,Y_j\rangle= 0,\qquad \langle y^i,y^j\rangle=0, \qquad
\langle y^i,Y_j\rangle= \delta^i_j,\quad \forall i,j  
\label{age}
\ee
and the ``double" vector space 
$\mathfrak{a}=\mathfrak{g} \oplus \mathfrak{g}^*$ can be endowed with a Lie algebra
structure, the  so-called Drinfel'd double~\cite{Drinfelda},  given by
\be
[Y_i,Y_j]= c^k_{ij}Y_k, \qquad  
[y^i,y^j]= 0, \qquad
[y^i,Y_j]= c^i_{jk}y^k   . \label{agd}
\ee
The Lie group $D(G)$ with tangent Lie bialgebra  $\mathfrak{a}=\mbox{Lie}(D(G))$ is the Drinfel'd double Lie group associated to the trivial Lie bialgebra $(\mathfrak{g},\delta_0)$. By construction, it is a semidirect product Lie group $G\ltimes \mathfrak g^*$.
Moreover, the double Lie algebra $D ({\mathfrak{g}})\equiv \mathfrak{a}$  can be endowed with a
(quasi-triangular) Lie bialgebra structure $(D ({\mathfrak{g}}),\delta_{D})$  that is defined 
by the ``canonical" classical
$r$-matrix
\be
r=\sum_i{y^i\otimes Y_i}
\label{rcanon}
\ee
or, equivalently, by its skew-symmetric counterpart 
$ r'=\frac12 \sum_i{y^i\wedge Y_i}  
\label{rmat}
$
via the  
coboundary relation
\be
\delta_{D}(X)=[ X \otimes 1 +1\otimes X ,r'],
\qquad
\forall X\in D ({\mathfrak{g}}).
\label{cob}
\ee
The cocommutator $\delta_{D}$  derived from (\ref{cob}) is then given by
\be
\delta_{D}(y^i)=\tfrac12 \,c^i_{jk}\,y^j\wedge y^k ,\qquad
\delta_{D}(Y_i)=0 .
\label{codob}
\ee
This means that the trivial Lie bialgebra structure $(\mathfrak{g},\delta_0)$~\eqref{lib} associated to the non-deformed Hopf algebra $(U(\mathfrak{g}),\Delta_0)$ induces  a unique non-trivial quantum deformation of the  double Lie algebra $D({\mathfrak{g}})$ whose first-order deformation in the coproduct is given by~\eqref{codob}. Therefore, in this deformation of $U(D({\mathfrak{g}}))$ the subalgebra  of $ D({\mathfrak{g}})$ generated by $\mathfrak g$ will be primitive and the coproduct of the subalgebra generated by $\mathfrak g^*$  (which is  abelian) contains all information about the deformation.

In the corresponding quantum group with generators $\{\hat y_i, \hat Y_j \}$, 
the first-order relations for the quantum coordinates on $D(G)$ would be given by the dual of $\delta_D$~\eqref{codob}. This means that the only non-vanishing relations for the  coordinates  operators will be given ---up to higher-order terms--- by
\be
[\hat y_i, \hat  y_j]= \frac12 \, c_{ij}^k \, \hat y_k.
\label{ncgen}
\ee
This is a  general construction that yields a non-commutative subset of local coordinates on the quantum double group whose commutation rules are just isomorphic to the ones given by the initial Lie algebra $\mathfrak{g}$.
In other words, any finite-dimensional Lie algebra $\mathfrak{g}$ induces a quantum deformation on the semidirect product Lie group   $D(G)=G\ltimes \mathfrak g^*$ in which a subset of non-commutative coordinates have commutation rules isomorphic to $\mathfrak{g}$. This is a canonical way to construct non-commutative spaces of Lie algebraic type with prescribed commutation rules. 

Now it is worth stressing that for a quantum deformation $(U_\m(\mathfrak{g}),\Delta_\m)$, the cocommutator $\delta_\m$ is no longer trivial ($f\neq 0$ in~\eqref{cocof}) and the Drinfel'd double Lie algebra is given by
\begin{align}
[Y_i,Y_j]= c^k_{ij}Y_k, \qquad  
[y^i,y^j]= f^{ij}_k y^k, \qquad
[y^i,Y_j]= c^i_{jk}y^k- f^{ik}_j Y_k  ,
\label{agdoub}
\end{align}
which means that the semidirect product structure is lost. As a consequence, the corresponding Drinfel'd double non-commutative spacetime will be a deformation of~\eqref{ncgen} with a deformation parameter related to $\eta$.


\section{The (2+1) Poincar\'e  Lie algebra   as a Drinfel'd double}

We will now illustrate the construction in the preceding section with the example of the Lie algebra $sl(2,\mathbb{R})$. Consider a basis of $sl(2,\mathbb{R})$ in which the Lie bracket takes the form
\be
 [Y_0,Y_1]= 2 Y_1 ,
\qquad 
  [Y_0,Y_2]=  -2 Y_2 ,
\qquad 
  [Y_1,Y_2]= Y_0 ,
\ee  
and the universal enveloping algebra  with its primitive (non-deformed) coproduct~\eqref{delta0}.
This corresponds to a vanishing  cocommutator map  $\delta_0(Y)=0$, and  the Drinfel'd double Lie algebra $D(sl(2,\mathbb{R}))$ is given by the relations~\eqref{agd}, namely
\begin{align} 
\begin{array}{lll}
  [Y_0,Y_1]= 2 Y_1 ,&
\qquad 
  [Y_0,Y_2]=  -2 Y_2 ,&
\qquad 
  [Y_1,Y_2]= Y_0,
\\[2pt]
 [y^0,y^1]= 0 ,&
\qquad 
[y^0,y^2]=0, &
\qquad
 [y^1,y^2]=0 ,
 \\[2pt]
[y^0,Y_0]=0, &
\qquad 
 [y^0,Y_1]=y^2 ,&
\qquad 
[y^0,Y_2]=-y^1,
\\[2pt]
[y^1,Y_0]=2 y^1,&
\qquad 
 [y^1,Y_1]=-2 y^0 ,&
\qquad  
[y^1,Y_2]=0,
\\[2pt]
[y^2,Y_0]=- 2 y^2 ,&
\qquad 
[y^2,Y_1]= 0, &
\qquad 
[y^2,Y_2]=2 y^0.
 \end{array}
\label{pdoub}
\end{align}
This is essentially the Drinfel'd double proposed in~\cite{Bais} as the algebraic structure providing the non-commutative geometry for (2+1) Lorentzian quantum gravity with vanishing cosmological constant.
Now, following the approach in~\cite{BHMplb,BHMCQG} we can identify this Lie algebra with the (2+1) Poincar\'e algebra through the following change of basis  
\begin{align} 
\begin{array}{lll}
J_0=-\frac12 (Y_1 -Y_2) ,& \qquad
J_1=\frac12  Y_0 , & \qquad
 J_2=\frac12 (Y_1 +Y_2)   , \\[2pt]
P_0=y^1 - y^2 , &\qquad
 P_1=2 y^0 ,&  \qquad
 P_2=y^1 + y^2.
  \end{array}
\label{csbasisp}
\end{align}
 It is immediate to check that the resulting Lie bracket is the one of the  Poincar\'e algebra $p(2+1)\equiv iso(2,1)$ in (2+1) dimensions
\begin{align} 
\begin{array}{lll}
 [J_0,J_1]=J_2, & \quad [J_0,J_2]=-J_1,  & \quad  [J_1, J_2]=  -J_0,  \\[2pt]
[J_0,P_0]=0 ,& \quad [J_0,P_1]=P_2 ,& \quad [J_0, P_2]=-P_1,\\[2pt]
[J_1,P_0]=-P_2 ,& \quad [J_1,P_1]=0 ,& \quad [J_1, P_2]=-  P_0,\\[2pt]
[J_2,P_0]=P_1 ,& \quad[J_2,P_1]=P_0 ,& \quad[J_2, P_2]=0,\\[2pt]
[P_0,P_1]=  0, & \quad[P_0, P_2]=0  ,& \quad  [P_1,P_2]= 0   ,
 \end{array}
\label{aa}
\end{align}
where $P_a$ and $J_a$ ($a=0,1,2$) are, respectively, the generators of translations and Lorentz trans\-formations.
The inverse change of basis reads
\begin{align} 
\begin{array}{lll}
  Y_0=2 J_1, & \qquad
 Y_1=-J_0+J_2 ,&  \qquad
 Y_2=J_0+J_2   , \\[2pt]
y^0=\frac12 P_1, & \qquad
 y^1=\frac12 (P_0 +P_2), &  \qquad
 y^2=\frac12 (-P_0 +P_2) ,
  \end{array}
\label{csbasisinv3}
\end{align}
 and in terms of the new basis the pairing~\eqref{age} is given by
\be
  \langle J_a,P_b\rangle=g_{ab},\qquad \langle J_a,J_b\rangle=\langle P_a,P_b\rangle=0,
\ee
with $g_{ab}=(-1,1,1)$. The fact that $p(2+1)$ can be interpreted as  the double Lie algebra $D(sl(2,\mathbb{R}))$ yields a canonical quasi-triangular Lie bialgebra structure on $p(2+1)$ that is generated by the classical $r$-matrix
$$
r=\sum{y^i \otimes Y_i}=-P_0\otimes J_0 + P_1\otimes J_1 + P_2\otimes J_2.
$$
By taking into account that the two quadratic Casimirs of $p(2+1)$  are given by
\be
C_1=-P_0^2+P_1^2+P_2^2,\qquad
C_2=-J_0 P_0 +J_1 P_1 +J_2 P_2 ,
\ee
we see that the symmetric part of $r$ is one half  the tensorised Casimir $C_2$.  This allows one to fully skew-symmetrize the $r$-matrix and yields
\be
r'= \tfrac{1}{2} (J_0\wedge P_0+
P_1\wedge J_1 + P_2\wedge J_2).
\ee
It is directly apparent from the structure of this $r$-matrix that it  generates a quantum deformation of $p(2+1)$, which is a superposition of three non-commuting twists. 
In order to analyse this deformation in more depth, we introduce a new quantum deformation parameter $\xi$ as a global multiplicative factor for the classical $r$-matrix that generates the deformation:
\be
r_\xi\equiv \xi r' = \tfrac{\xi}{2} (J_0\wedge P_0+
P_1\wedge J_1 + P_2\wedge J_2).
\label{rpoincare}
\ee
The cocommutator induced by $r_\xi$ is given by (\ref{cob}) and reads
\bea
&&\delta_\xi(J_0)=\delta_\xi(J_1)=\delta_\xi(J_2)=0 ,\nonumber\\[2pt]
 &&\delta_\xi(P_0)= \xi P_1\wedge P_2 ,\qquad \delta_\xi(P_1)=\xi \, P_0\wedge P_2 ,\qquad \delta_\xi(P_2)=\xi  P_1\wedge P_0 ,
\label{momenta}
\eea
and defines the first-order in $\xi$ of the full quantum coproduct.
Therefore, equation~\eqref{momenta} implies that the corresponding full quantum deformation $U_\xi(D(sl(2,\mathbb{R})))\simeq U_\xi(p(2+1))$ (the quantum double) has a non-deformed Lorentz sector
\be
\Delta_\xi(J_a)=\Delta_0(J_a)= J_a\otimes 1+1\otimes J_a, \qquad a=0,1,2
\ee
and the full deformation is concentrated in the subalgebra of translations, which is an Abelian subalgebra before deformation. Consequently, both the Lorentz and the translation sectors will be Hopf subalgebras and no modification of the commutation rules~\eqref{aa} is expected.

On the other hand, if we introduce the coordinate functions $(\hat x_a,\hat \theta_a )$   that are dual to the generators $(P_a,J_a)$ $(a=0,1,2)$ by setting
\begin{align}\label{dualdef}
\langle\hat x_a, P_b\rangle=\delta_{ab},\qquad \langle\hat x_a, J_b\rangle=0,\qquad \langle\hat \theta_a, P_b\rangle=0,\qquad \langle\hat \theta_a, J_b\rangle=\delta_{ab} ,
\end{align}
the quantum group dual to $U_\xi(D(sl(2,\mathbb{R})))$ would be characterised in first-order only by  non-vanishing relations obtained from dualising~\eqref{momenta}, namely
\begin{align}
 \label{ads}
[\hat x_0, \hat  x_1]=- \xi \,  \hat x_2 , \qquad [\hat x_0, \hat x_2]=\xi \,  \hat x_1,  \qquad [\hat x_1, \hat x_2]= \xi \,  \hat x_0 ,
\end{align}
which means that, up to higher-order terms in the quantum coordinates, the non-commutative spacetime linked to this quantum double is  just the (1+1) AdS Lie algebra $so(2,1)$, as proposed  in~\cite{Bais}.

In principle, when all orders in the quantum coordinates are considered, the spacetime~\eqref{ads} could exhibit further non-linear contributions. The easiest way to address this question is to construct the Poisson bracket that defines the unique PL structure on the Poincar\'e group  $P(2+1)=ISO(2,1)$ induced by the classical $r$-matrix $r_\xi$ (\ref{rpoincare}).  The quantisation of this PL structure will then provide the full non-commutative spacetime associated to this quantum double. As we will show in section 5,  in this case the  PL structure is  given by the Poisson analogues of the above relations, which just coincide with (\ref{ads}), namely
\begin{align}
 \label{adspois}
 \{ x_0,  x_1\}= -\xi  \, x_2 , \qquad \{ x_0, x_2\}= \xi  \,  x_1,  \qquad \{x_1, x_2\}= \xi  \, x_0,
\end{align}
and the remaining Poisson brackets vanish. The relations (\ref{ads}) therefore define the non-com\-mu\-tative Minkowski spacetime ${\bf M}_\xi^{2+1}$, and they are compatible with the coproduct $\Delta_\xi$. The latter  is given by the multiplication law of the corresponding quantum  Poincar\'e group which, in this case, has non-commutative group parameters only in the quantum translations sector. 

This construction is the one underlying  all  previous investigations of such quantum double spacetimes (see~\cite{MW, BatistaMajid, Majidtime, Noui}). All of them are Lie algebraic spacetimes, and the representation theory of the corresponding algebra ($so(2,1)$ in the ${\bf M}_\xi^{2+1}$ case) characterises their physics properties. On the other hand, we recall that such Lie algebraic deformation of (2+1) Minkowski space was obtained by twisting the (2+1) Poincar\'e algebra in~\cite{lukiworo}, without making use of the underlying Drinfel'd double structure.


\section{The  (2+1) AdS  Lie algebra as a Drinfel'd double}

In the remainder of the paper, we show  how the cosmological constant can be  introduced into the previous construction by considering the standard quantum deformation of $sl(2,\mathbb{R})$ as the starting point of the construction. In fact, as shown in~\cite{BHMplb, BHMCQG}, the Drinfel'd double group associated to this deformation is just the isometry group of AdS in (2+1) dimensions.

Recall first that the so-called
standard (or Drinfel'd--Jimbo~\cite{Drinfelda,Jimbo}) quantum deformation of $sl(2,\mathbb{R})$ is the Hopf algebra defined by
\be
[Y_0,Y_1]= 2 Y_1 ,\qquad [Y_0,Y_2]=- 2 Y_2 ,  \qquad  [Y_1 ,Y_2]=\frac{\sinh (\m Y_0)}{\m},
\ee
\begin{align} 
\begin{array}{l}
  \Delta_\m( Y_0) =Y_0 \otimes 1 + 1\otimes Y_0 , \\[2pt]
 \Delta_\m (Y_{1}) =Y_{1} \otimes {\rm e}^{\frac \m 2 Y_0} + {\rm e}^{-\frac \m 2 Y_0} \otimes Y_{1}  , \\[2pt]
 \Delta_\m (Y_{2}) =Y_{2} \otimes {\rm e}^{\frac \m 2 Y_0} + {\rm e}^{-\frac \m 2 Y_0} \otimes Y_{2}  .
 \end{array}
\end{align}
In the following, we denote this Hopf algebra by  $sl_\m(2,\mathbb{R})$, where initially $\m$ is a {\em real} deformation parameter (and $q={\rm e}^\m$). The non-trivial Lie bialgebra structure associated to this deformation is given by
\be
\delta_\m(Y_0)=0,\qquad  
\delta_\m(Y_{1})=\tfrac{\m}{2} \,Y_{1}\wedge Y_0 ,
\qquad
\delta_\m(Y_{2})=\tfrac{\m}{2} \,Y_{2}\wedge Y_0.
\label{cocostandard}
\ee
This Lie bialgebra is generated by the classical $r$-matrix $r=\tfrac{\eta}{2}\,Y_1\wedge Y_2$ via the coboundary condition~\eqref{cob}.
In this case, the double Lie algebra $D(sl_\m(2,\mathbb{R}))$ is obtained from~\eqref{agdoub}:
\begin{align} 
\begin{array}{lll}
  [Y_0,Y_1]= 2 Y_1 ,&
\qquad 
  [Y_0,Y_2]=  -2 Y_2 ,&
\qquad 
  [Y_1,Y_2]= Y_0,
\\[2pt]
 [y^0,y^1]= -\frac{\m}{2} y^1 ,&
\qquad 
[y^0,y^2]=-\frac{\m}{2} y^2 ,&
\qquad
 [y^1,y^2]=0,
\\[2pt]
[y^0,Y_0]=0 ,&
\qquad 
 [y^0,Y_1]=y^2+\frac{\m}{2} Y_1 ,&
\qquad 
[y^0,Y_2]=-y^1+\frac{\m}{2} Y_2,
\\[2pt]
[y^1,Y_0]=2 y^1,&
\qquad 
 [y^1,Y_1]=-2 y^0-\frac{\m}{2} Y_0, &
\qquad  
[y^1,Y_2]=0,
\\[2pt]
[y^2,Y_0]=- 2 y^2 ,&
\qquad 
[y^2,Y_1]= 0, &
\qquad 
[y^2,Y_2]=2 y^0-\frac{\m}{2} Y_0 .
 \end{array}
\label{pdoubads}
\end{align}

As shown in~\cite{BHMplb, BHMCQG}, this   Lie algebra is  isomorphic to the isometry algebra   of  the (2+1) AdS space. In terms of the alternative basis 
 \begin{align}
&J_0=-\tfrac12 (Y_1 -Y_2) , & 
&J_1=\tfrac12 \,Y_0 ,&  
&J_2=\tfrac12 (Y_1 +Y_2)   , \label{csbasis3}\\
&P_0=-\tfrac{\m}{2}\, (Y_1 +Y_2) +(y^1 - y^2),& 
&P_1=2\,y^0,&  
&P_2=\tfrac{\m}{2}  (Y_1 -Y_2) +(y^1 + y^2) ,
\nonumber
\end{align}
 the Lie bracket reads
 \begin{align} 
\begin{array}{lll}
 [J_0,J_1]=J_2, & \quad [J_0,J_2]=-J_1,  & \quad  [J_1, J_2]=  -J_0,  \\[2pt]
[J_0,P_0]=0 ,& \quad [J_0,P_1]=P_2 ,& \quad [J_0, P_2]=-P_1,\\[2pt]
[J_1,P_0]=-P_2 ,& \quad [J_1,P_1]=0 ,& \quad [J_1, P_2]=-  P_0,\\[2pt]
[J_2,P_0]=P_1 ,& \quad[J_2,P_1]=P_0 ,& \quad[J_2, P_2]=0,\\[2pt]
[P_0,P_1]=  {\m^2}  J_2, & \quad[P_0, P_2]=-{\m^2} J_1  ,& \quad  [P_1,P_2]= -{\m^2  }  J_0   .
 \end{array}
\label{adseta}
\end{align}
Following~\cite{BHMCQG}, we realise that~\eqref{adseta} is the Lie bracket of  $so(2,2)$, and the deformation parameter $\m$ is   directly related to  the (negative) cosmological constant $\Lambda$ through  
\be
  \Lambda= -\m^2  .
\label{ba}
\ee
In fact, the Lie brackets \eqref{aa} and \eqref{adseta} are precisely the Lie brackets from \cite{Witten1}, which allow one to express the Lie algebras $p(2+1)\equiv iso(2,1)$, $so(2,2)$ and $sl(2,\mathbb C)\simeq so(3,1)$ of the isometry groups of (2+1)-dimensional Minkowski, AdS  and dS spaces in terms of a common basis, such that the cosmological constant appears as a structure constant.

The AdS quadratic Casimirs are given by
   \be 
{  C}_ 1= -P_0^2 +P_1^2+P_2^2+\m^2 (  -J_0^2+J_1^2+J_2^2)  , \qquad
{  C}_2=  -J_0 P_0+ J_1P_1+J_2P_2, 
\label{bc}
\ee  
and the canonical pairing of the Drinfel'd double  reads
\begin{align}
\label{jppair}
& \langle J_0, P_0\rangle=-1 ,\qquad \langle J_1, P_1\rangle=1 ,\qquad \langle J_2, P_2\rangle=1 , \\
&\langle J_a, J_b\rangle=\langle P_a, P_b\rangle=0, \qquad \langle J_a, P_b\rangle=0\quad  \text{for}\ a\neq b,\quad a,b=0,1,2 ,\nonumber 
\end{align}
which is exactly the appropriate pairing for the CS  formulation of (2+1) gravity on the constant curvature space whose isometries are given by~\eqref{adseta}.

By inverting~\eqref{csbasis3} one finds that the canonical classical $r$-matrix~\eqref{rcanon} inherited  from the Drinfel'd double structure reads 
 \begin{align}
r= {\m}  J_0\wedge   J_2+ \left( -  P_0\oo   J_0+  P_1\oo   J_1+  P_2\oo   J_2\right) ,
\label{rCS}
\end{align}
and its fully skew-symmetric counterpart is obtained by subtracting the tensorised Casimir ${ C}_2$ (see~\cite{BHMCQG} for details)
 \begin{align}
 r'= {\m}  J_0\wedge   J_2+\tfrac 12 \left( -  P_0\wedge   J_0+  P_1\wedge   J_1+  P_2\wedge   J_2\right) .
 \label{rads}
\end{align}
Again, we will multiply $r'$ by the quantum double deformation parameter $\xi$, in such a way that
the classical $r$-matrix $r_\xi\equiv \xi r'$ defines a quantum deformation $U_\xi(D(sl_\m(2,\mathbb{R})))\simeq U_\xi(\mbox{AdS})$.
Moreover,  $r_\xi$ defines the unique PL  structure on the AdS group manifold  that is associated to the previous double structure. As we will see in the sequel, once this Poisson--Hopf algebra is obtained in appropriate coordinates, its quantisation provides the quantum AdS group dual to $U_\xi(D(sl_\m(2,\mathbb{R})))$, and the non-commutative AdS spacetime ${\bf AdS}_\xi^{2+1}$ will arise as the quantisation of the PL brackets among the space and time coordinates.

Explicitly, the  cocommutator generated by 
$r_\xi  $ reads
\begin{align}\label{cocomm}
&\delta_\xi(  J_0)=\m\xx     J_1\wedge   J_0,\qquad \delta_\xi(  J_1)=0,\qquad \delta_\xi(  J_2)=\m\xx    J_1\wedge   J_2 , &&\nonumber\\
&\delta_\xi(  P_0)=\xx \left(    P_1\wedge  P_2 +\m     P_1\wedge   J_0+\m^2  J_2\wedge   J_1 \right) ,& &\\
&\delta_\xi(  P_1)=\xx \left(       P_0\wedge   P_2+\m       P_0\wedge   J_0 - \m    P_2\wedge  J_2+\m^2      J_2\wedge   J_0\right) ,& &\nonumber\\
&\delta_\xi(  P_2)=\xx \left(        P_1\wedge  P_0   +\m       P_1\wedge   J_2 +\m^2       J_0\wedge  J_1\right) ,& &\nonumber
\end{align}
which gives  the first-order term  $\Delta_1$ of the full quantum coproduct in $U_\xi(D(sl_\m(2,\mathbb{R})))$. Note that the zero cosmological constant limit is obtained by taking $\m\to 0$ in all the above expressions, and leads to the   (2+1)-Poincar\'e quantum double with classical $r$-matrix~\eqref{rpoincare}. 

In terms of the dual basis $(\hat x_a,\hat \vv_a )$ $(a=0,1,2)$ defined by (\ref{dualdef}),
we find from~\eqref{cocomm} that  the first-order dual Lie brackets among the spacetime coordinates are given by
\begin{align}
 \label{xxbrack3}
[\hat x_0, \hat  x_1]=- \xx   \hat x_2 ,\qquad [\hat x_0, \hat x_2]=\xx  \hat x_1, \qquad [\hat x_1, \hat x_2]= \xx  \hat x_0.
\end{align}
However, the additional terms in~\eqref{cocomm} that appear due to the non-vanishing cosmological constant $\m$ give rise to the first-order non-commutative relations between the quantum spacetime and Lorentz parameters:
\begin{align}
&[\hat \vv_0, \hat \vv_1]= -\m   \xx  (\hat \vv_0-\m   \hat x_2  )  ,&  &[\hat \vv_0, \hat \vv_2]=- \m^2 \xx   \hat x_1, & &[\hat \vv_1, \hat \vv_2]=\m     \xx   ( \hat \vv_2- {\m}    \hat x_0  ),\nonumber
\\
&[\hat \vv_0, \hat x_0]=-\m  \xx   \hat x_1 ,& &[\hat \vv_0, \hat x_1]= -\m   \xx  \hat x_0,& &[\hat \vv_0, \hat x_2]=0 ,\nonumber\\
&[\hat \vv_1, \hat x_0]=0 ,& &[\hat \vv_1, \hat x_1]=0 ,& &[\hat \vv_1, \hat x_2]=0,\label{vxbrack2}\\
&[\hat \vv_2, \hat x_0]=0 ,& &[\hat \vv_2, \hat x_1]=-\m    \xx  \hat x_2 ,& &[\hat \vv_2, \hat x_2]=\m   \xx   \hat x_1\nonumber.
\end{align}

Therefore, up to higher-order corrections in the quantum spacetime coordinates, the non-commutative   space ${\bf AdS}_\xi^{2+1}$ is again isomorphic to $sl(2,\mathbb R)\simeq so(2,1)$, and coincides with the quantum Minkowski space~\eqref{ads} obtained in the previous section. Nevertheless, modifications are expected to arise in~\eqref{xxbrack3} when higher-orders in terms of the quantum spacetime coordinates are considered, since in this case $\m\neq 0$. To obtain such higher-order terms explicitly, one must construct the full AdS quantum algebra $U_\xi(D(sl_\m(2,\mathbb{R})))$  (recall that~\eqref{cocomm} gives only the first-order deformation of the coproduct) and, afterwards, compute its dual Hopf algebra. However, this lengthy procedure can be circumvented by computing directly the  PL  brackets associated to the $r$-matrix $r_\xi\equiv \xi r'$  (\ref{rads})  in terms of the {\em classical} AdS coordinates $(x_a,\vv_a)$ $(a=0,1,2)$, since the quantisation of this PL  algebra  will provide the all-orders AdS non-commutative spacetime in terms of the {\em quantum} coordinates $(\hat x_a,\hat \vv_a)$.


 \section{The  (2+1)  AdS PL  group and non-commutative spacetime}
 
 It is well known that the action of the isometry  group    $SO(2,2)$ on its homogeneous space
 $$
 {\bf AdS}^{2+1}=SO(2,2)/SO(2,1),\qquad SO(2,1)=\langle J_0,J_1,J_2\rangle
 $$
   is nonlinear. However, a linear $SO(2,2)$ action can be obtained by considering the
vector representation of the Lie group   which makes use of an
 ambient space with  an ``extra" dimension. In particular, the $ 4\times  4$ real
matrix representation of $so(2,2)$  with Lie brackets \eqref{adseta} is given by
\begin{align}
&   P_0=\left(\begin{array}{cccc}
0&- \m^2&0&0\cr 
1&0&0&0\cr 
0&0&0&0\cr 
0&0&0&0
\end{array}\right) ,  \quad 
  P_1=\left(\begin{array}{cccc}
0&0& \m^2 &0\cr 
0&0&0&0\cr 
1&0&0&0\cr 
0&0&0&0
\end{array}\right) ,  \quad 
  P_2=\left(\begin{array}{cccc}
0&0&0& \m^2 \cr 
0&0&0&0\cr 
0&0&0&0\cr 
1&0&0&0
\end{array}\right)  , \nonumber\\[4pt]
&   J_0=\left(\begin{array}{cccc}
0&0&0&0\cr 
0&0&0&0\cr 
0&0&0&-1\cr 
0&0&1&0
\end{array}\right) ,   \quad 
  J_1=\left(\begin{array}{cccc}
0&0&0&0\cr 
0&0&0&-1\cr 
0&0&0&0\cr 
0&-1&0&0
\end{array}\right)  , \quad 
  J_2=\left(\begin{array}{cccc}
0&0&0&0\cr 
0&0&1&0\cr 
0&1&0&0\cr 
0&0&0&0
\end{array}\right)  ,
\label{bd}
\end{align}
and fulfils
\be
Y^{T}\mathbb I_{\m} +\mathbb I_{\m} Y=0  ,\quad Y\in so (2,2), \quad
\mathbb I_{\m}={\rm diag}\,(1, \m^2,- \m^2 ,- \m^2  ) .
\label{be}
\ee
The exponential of (\ref{bd}) leads
to the vector representation of  
$SO(2,2)$ as a Lie group of  matrices which acts linearly, via matrix multiplication, on 
 a 4-dimensional  space with  ambient or Weierstrass
coordinates 
$(\s_3,\s_0,\s_1,\s_2)$. By definition, any element $G\in   SO(2,2)$ satisfies the relation
$G^T\mathbb I_{\m} G=\mathbb I_{\m}$.   
Note that this realisation includes explicitly the cosmological constant parameter $\m$, and the   one-parameter subgroups of  $SO(2,2)$ obtained from  (\ref{bd}) are, for instance, 
$$
{\rm e}^{x_0   P_0}=\left(\begin{array}{cccc}
\cos  \m   x_0&- \m \sin   \m x_0&0&0\cr 
\displaystyle \frac {\sin   \m x_0} { \m} &\cos   \m x_0&0&0\cr 
0&0&1&0\cr 
0&0&0&1
\end{array}\right),
\quad
{\rm e}^{x_1   P_1}=\left(\begin{array}{cccc}
\cosh   \m   x_1&0& \m   \sinh  \m   x_1&0\cr 
0&1&0&0\cr 
\displaystyle \frac {\sinh  \m   x_1} { \m  } &0&\cosh  \m   x_1&0\cr 
0&0&0&1
\end{array}\right).
$$
In this vector model, the 3-dimensional space ${\bf AdS}^{2+1}$  is
identified with the orbit containing the origin of the 4-dimensional space  
$O= (1,0,0,0)$, which is  contained in the pseudosphere $\Sigma_{\m}$ provided by
${\mathbb I}_{\m}$:
\be
\Sigma_{\m}:\  \s_3^2 + \m^2 (\s_0^2 -    \s_1^2-  \s_2^2)=1 .
\label{Drinfel'd double}
\ee
 Note that any element of the Lorentz subgroup $SO (2,1)=\langle J_0,J_1,J_2\rangle $ leaves the origin $O$ invariant.
The  metric on ${\bf AdS}^{2+1}$ comes 
from the flat ambient metric  divided by $\m^2$ (the sectional curvature) and
restricted to the above pseudosphere constraint:
\be
\Drinfel'd double s^2  =   \left. \frac{1}{ \m^2}
\left(\Drinfel'd double \s_3^2+ \m^2 (\Drinfel'd double\s_0^2 -    \Drinfel'd double\s_1^2-   \Drinfel'd double\s_2^2)\right)
\right|_{\Sigma_{\m }} .
\label{de}
\ee
 In particular, let us      consider a generalisation of the Cartesian coordinates to curved spaces known as  ``geodesic
parallel coordinates" $x_a$~\cite{conf} which are the classical counterpart of the quantum coordinates $\hat x_a$ (\ref{dualdef}). They are
defined     through
the following action of the translation subgroups on the origin $O$
\be
(\s_3,\s_0,\s_1,\s_2 )(x_0,x_1,x_2)=\exp(x_0 P_0)\exp(x_1 P_1)\exp(x_2 P_2)  O ,
\label{df}
\ee
which yields
\be
\begin{array}{l} 
\s_3=\cos \m x_0 \cosh \m  x_1 \cosh \m  x_2 ,\\ 
\displaystyle {\s_0=\frac {\sin \m x_0}{\m}  \cosh \m  x_1 \cosh \m 
x_2 },\\[8pt]  
\displaystyle {\s_1=\frac {\sinh \m  x_1 }{\m }\,   \cosh \m  x_2 },\\[8pt] 
\displaystyle {\s_2=\frac { \sinh \m  x_2}{\m }  . }
\end{array}
\label{dg}
\ee
In terms of these coordinates, the   metric (\ref{de}) reads
\be
\Drinfel'd double s^2 =\cosh^2(\m  x_1) \cosh^2(\m  x_2)\Drinfel'd double x_0^2-  \cosh^2(\m 
x_2)\Drinfel'd double x_1^2-  \Drinfel'd double x_2^2 .
\label{dh}
\ee
 If $\eta\to 0$, the
parametrisation (\ref{dg}) gives the  flat Cartesian coordinates 
$\s_3=1,  \s_a= x_a$, and the  metric (\ref{dh}) reduces to 
$\Drinfel'd double s^2 = \Drinfel'd double x_0^2-  \Drinfel'd double
x_1^2- \Drinfel'd double x_2^2$, which is the metric  of  the  classical Minkowski space ${\bf M}^{2+1}$.

Consider now the  $4\times 4$ matrix element of the group
$SO(2,2)$ obtained through
\be
T=\exp(x_0 P_0)\exp(x_1 P_1)\exp(x_2 P_2) \exp(\vv_2 J_2)\exp(\vv_1 J_1)
\exp(\vv_0 J_0)  
\label{ga}
\ee
where the group coordinates are the ones  defined above.
The PL  brackets associated to a given  
classical $r$-matrix $r'= 
r^{ij}X_i\wedge X_j$  and defined on
the algebra of smooth functions $C^\infty(SO(2,2))$, are obtained from the Sklyanin bracket
\cite{Drib}: 
\be
\{f,g\}=  r^{ij}(X_i^Lf\, X_j^L g -
X_i^Rf\, X_j^R g),  \qquad f,g\in   C^\infty(SO(2,2)).
\label{gb}
\ee
Thus, after computing from~\eqref{ga} the $SO(2,2)$ left- and right-invariant vector fields, $X^L$ and $X^R$, one obtains the 
PL brackets between the six  {commutative} group coordinates  $
 (x_a,  \vv_a ) $ $(a=0,1,2)$ associated to the
classical $r$-matrix    $r_\xi\equiv \xi r'$  \eqref{rads}.
The main point of interest are  the brackets defined by the $x_a$ group
coordinates, which  read
\be
\begin{aligned}
& \{x_0,x_1\} = -
\xi\frac{\tanh\m x_2 }{\m} 
\,\C,
\\
&
 \{x_0,x_2\} = \xi
 \frac{ \tanh\m x_1}{\m}\,
\C ,
\\
& \{x_1,x_2\} = \xi
\frac{\tan\m x_0}{\m}
\,\C , \qquad \text{where}\qquad\C(x_0,x_1)=\cos\m x_0(\cos\m x_0\cosh\m x_1+ \sinh\m x_1).
\label{qa}
\end{aligned}
\ee
The expressions \eqref{qa} for the Poisson brackets are surprisingly elegant and simple, and they involve the deformation parameter $\eta$ related to the cosmological constant in a symmetric way.  Note also that this Poisson structure is not symplectic. Its symplectic leaves are the level surfaces of  the function 
$$
C=\cos(\eta x_0)\cosh(\eta x_1)\cosh(\eta x_2)
$$
which Poisson commutes with all coordinate functions. By comparing this expression with \eqref{dg}, one finds that this Casimir function coincides with the ambient coordinate $\s_3$. In the limit $\eta\to 0$, this Casimir function becomes constant. Note, however, that the associated Casimir function $C'=2(1-C)/\m^2$ can be defined in such a way that it satisfies
$$
\lim_{\eta\to 0}\frac  {2(1-C)}{\eta^2}=x_0^2-x_1^2-x_2^2
$$
which is the quadratic Casimir function for the corresponding Poisson bracket in the Minkowski case, see equation \eqref{xxbrack3}.

The quantisation of the Poisson algebra~\eqref{qa} would be the quantum non-commutative spacetime with cosmological constant ${\bf AdS}_\xi^{2+1}$ that we are looking for. It becomes clear that the complete quantum spacetime
 ${\bf AdS}_\xi^{2+1}$  for $\m\ne 0$ is different from ${\bf M}_\xi^{2+1}$~\eqref{ads}, which is  obtained from it in the limit $\m\to0$,
 in full agreement with the first-order spacetime Lie brackets (\ref{xxbrack3}). In fact, if we consider the power series expansion of (\ref{qa})  in terms of the cosmological constant $\m$ we obtain
\be 
\begin{aligned}
& \{x_0,x_1\} =- \xi \, x_2 -  \m \xi \, x_1 x_2 + \m^2 \xi \left( x_0^2 x_2 - \tfrac{1}{2}\,x_1^2 x_2+ \tfrac{1}{3}\,x_2^3
\right )
+ o[\m^3] ,
\\
& \{x_0,x_2\} =  \xi x_1 +
  \m \xi x_1^2 - \m^2 \xi \left( x_0^2 x_1 - \tfrac{1}{6}\,x_1^3
\right )
+ o[\m^3] ,
\\ 
& \{x_1,x_2\} =  \xi x_0 +
  \m \xi x_0 x_1- \m^2 \xi \left( \tfrac{2}{3}\, x_0^3 - \tfrac{1}{2}\,x_1^2 x_0
\right )
+ o[\m^3] .
\end{aligned}
\ee
The remaining Poisson brackets between the classical coordinates can be straightforwardly computed from~\eqref{gb}. They are, in general, non-vanishing, which means that this quantum double is much more complicated than the expressions obtained in the limit  $\m\to 0$. The latter  are just given by the Poisson analogues of the relations~\eqref{xxbrack3}--\eqref{vxbrack2} that include the quantum Minkowski spacetime $ {\bf M}_\xi^{2+1}$. As a consequence, the quantisation of~\eqref{qa} seems to be quite complicated, since many ordering problems appear. Nevertheless, the ambient space variables $( \s_3,\s_a)$ (\ref{dg}) are  more accessible in this respect, since their  PL brackets turn out to be homogeneous quadratic brackets, namely
\be
\begin{aligned}
&\{ \s_0 ,\s_1\}=-\xi \s_2 (\s_3+\m \s_1),\qquad \{ \s_0 ,\s_2\}=\xi \s_1 (\s_3+\m \s_1),\\
& \{ \s_1 ,\s_2\}=\xi \s_0 (\s_3+\m \s_1),\qquad\;\;\; 
\{\s_3,\s_a\}=0,\quad a=0,1,2.
\end{aligned}
\label{qb}
\ee
In agreement with the remark after equation \eqref{qa}, the coordinate $\s_3$ turns out to be a Casimir function for the PL bracket. Note also that in  the limit of zero cosmological constant $\m\to 0$ we again have  $\s_3\to 1$ and $\s_a\to x_a$, such that the expressions in \eqref{qb}  reduce to the classical counterpart of (\ref{xxbrack3}).


 \section{Remarks and open problems}
 
The results of this article strongly suggest that the Lie algebraic non-commutative Minkowskian spacetimes   that were proposed in the literature have to be transformed into non-linear algebras when the cosmological constant does not vanish. The algebra~\eqref{qa} makes this assertion explicit, and the representation theory of its quantum version is a challenging open problem. Also, it is well known that non-commutative spacetimes with zero cosmological constant of the type in~\eqref{ads} are associated with curved momentum spaces for point particles that are related to certain group manifolds~\cite{Snyder, MW,Kowalski,KowalskiNowak, Freidel}. The generalization of this construction to the case with non-vanishing $\Lambda$ would have as a prerequisite the knowledge of the full quantum group generated by the $r$-matrix $r_\xi$, in order to apply the Heisenberg double construction~\cite{STS, LN}. Nevertheless, some insight into the complexity of the outcoming structure can be inferred from the results here presented:  such a momentum space structure with non-vanishing cosmological constant would have non-commuting momenta~\eqref{adseta}, the dispersion relation coming from the deformed analogue of the Casimir $C_1$ in~\eqref{bc} would include the Lorentz sector --which would also become quantum deformed, as it can be deduced from the first-order deformation given by the cocommutator~\eqref{cocomm}-- and the associated noncommutative spacetime~\eqref{ads} is not of Lie algebraic type. All these facts seem to indicate that this problem goes far beyond the framework on which both standard and novel approaches~\cite{relloc,AGP,Banburski} to curved momentum spaces are based on, and it deserves a separate study.

Moreover, besides the PL structure  presented here, the classical $r$-matrix $r_\xi$  can also be used to construct the so-called dual PL  and the Heisenberg double Poisson structure that are essential in the description of point particles on compact surfaces in the CS formulation of (2+1)-gravity \cite{BR,AGSI,AS, BNR, we2}. The effect of the deformation by the cosmological constant should therefore also be  studied in  these contexts, see also \cite{quat}. Finally, as it was explicitly shown in~\cite{BHMCQG}, there exist two other, non-equivalent,  realisations   of the isometry algebra of  the (2+1) AdS algebra as a Drinfel'd double. The corresponding non-commutative spacetimes can be constructed by following the  same approach as in this article. Work on these questions is in progress and will be presented elsewhere.


\section*{Acknowledgements}

This work was partially supported by the Spanish MICINN   under grant    MTM2010-18556 (with EU-FEDER support) and by the German DFG Emmy-Noether fellowship ME 3425/1-1.



\vfill
\eject

\end{document}